# INFLUENCE OF THE USER IMPORTANCE MEASURE ON THE GROUP EVOLUTION DISCOVERY

Stanisław SAGANOWSKI*, Piotr BRÓDKA*, Przemysław KAZIENKO*

**Abstract.** One of the most interesting topics in social network science are social groups, i.e. their extraction, dynamics and evolution. One year ago the method for group evolution discovery (GED) was introduced. The GED method during extraction process takes into account both the group members quality and quantity. The quality is reflected by user importance measure. In this paper the influence of different user importance measures on the results of the GED method is examined and presented. The results indicate that using global measures like social position (page rank) allows to achieve more precise results than using local measures like degree centrality or no measure at all.

## 1. Introduction and related work

In modern telecommunication systems, the relationships between users are very often discovered based on system logs, containing information on the elementary events - usually relating to service oriented system (message e-mail, phone call, etc.). Events are discrete, however the relationships used in further analysis of the social network are continuous. One aspect of the social network analysis is to investigate dynamics of a community, i.e., how particular group changes over time.

To deal with this problem several methods for tracking group evolution have been proposed: GraphScope [8], Chakrabarti method [4], FacetNet [5], Palla method [7], Asur method [1].This methods uses simple *overlapping measure* to calculate similarity between groups in successive timeframes. This means that only the number of users existing in both groups are taken into account to decide how similar investigated groups are. Such comparison ignores the structure of the social group, in particular the relations between users.

The GED method [3], to extract group evolution history utilize innovative *inclusion measure*, which considers not only the number of users but also their importance within

* Institute of Informatics, Wrocław University of Technology

a group. The user importance might be reflected by a number of user importance measures (also called metrics) like degree centrality, closeness centrality, betweenness centrality [9] or social position [6]. Comparison of user position measure can be found in [6] and [2], each measure has different properties and computational complexity and is designed to be used under different conditions. For example, degree centrality is a local measure and very easy to compute so it can be used in very large networks but produces big number of duplicates and since it is a local measure takes into account only first level neighbourhood. Betweenness centrality is a global measure which express the extent to which a given node controls the communication between two nodes, unfortunately it has huge computational complexity. In the middle we have measures like page rank or social position, which are global measures but calculated based on the positions of their neighbours so they have low computational complexity and produces very diverse results.

The GED method introduced in [3] was comprehensively analysed and compared with other methods for group evolution extraction. The experiments indicated that the method is very good, flexible and fast. Unfortunately, during experiments only social position measure was used so the method results may depend on the selected user importance measure and this issue has not been addressed by the authors before.

Thus, in this paper the influence of different measures on the results of group evolution discovery method is analysed to indicate whether it can be used witch variety of user importance measures or only with social position measure.

The rest of this paper is organized as follows: chapter 2 describes key concepts necessary for understanding the mechanics behind the GED method and the GED method itself. Chapter 3 briefly introduced user importance measures used in evaluation. Chapter 4 contains the experiments and is followed by conclusions (chapter 5) .

## 2. Group Evolution Discovery

Before the method can be presented, it is necessary to describe a few concepts related to social networks.

### 2.1. Temporal Social Network

Temporal social network TSN is a list of following timeframes (time windows) T. Each timeframe is in fact one social network SN(V,E) where V – is a set of vertices and E is a set of directed edges $<x,y>$: $x,y \in V$, $x \neq y$

$$
\begin{aligned}
& TSN = <T_1, T_2, ...., T_m>, \quad m \in N \\
& T_i = SN_i(V_i, E_i), \quad i = 1, 2, ..., m \\
& E_i = <x, y>: x, y \in V_i, x \neq y \quad i = 1, 2, ..., m
\end{aligned}
\quad (1)
$$

## 2.2. Group evolution

Evolution of particular social community can be represented as a sequence of events (changes) following each other in the successive time windows (timeframes) within the temporal social network. Possible events in social group evolution are:

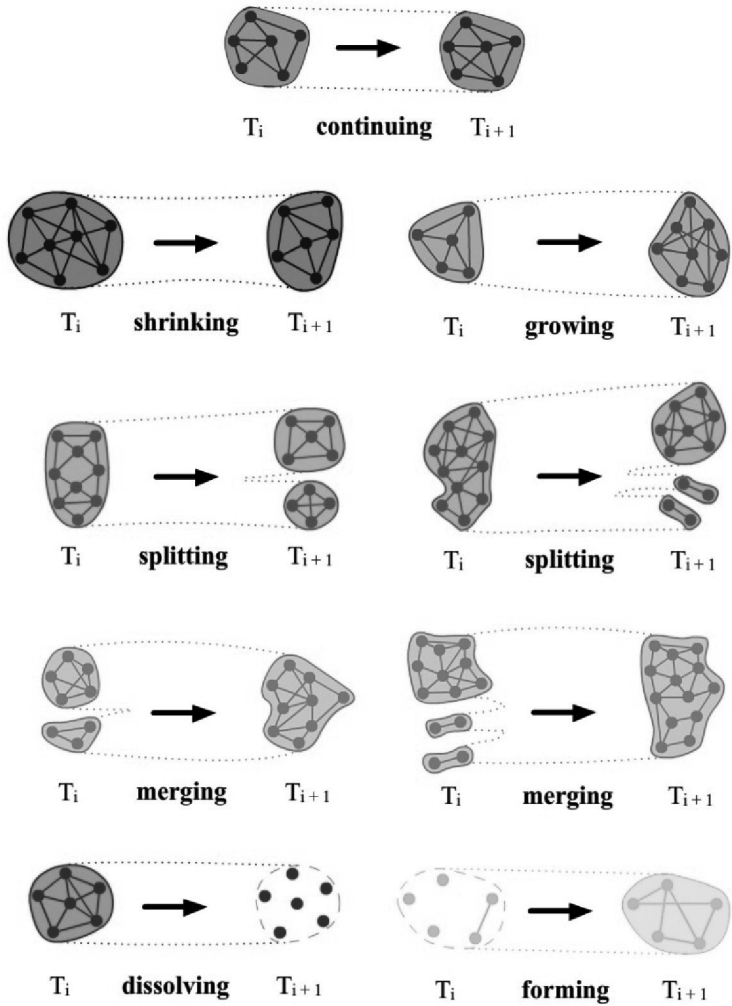

**Figure 1.    The events in group evolution.**

1.    Continuing (stagnation) – the group continue its existence when two groups in the consecutive time windows are identical or when two groups differ only by few nodes but their size remains the same.

2.    Shrinking – the group shrinks when some nodes has left the group, making its size smaller than in the previous time window. Group can shrink slightly, i.e. by a few nodes or greatly losing most of its members.

3. Growing (opposite to shrinking) – the group grows when some new nodes have joined the group, making its size bigger than in the previous time window. A group can grow slightly as well as significantly, doubling or even tripling its size.

4. Splitting – the group splits into two or more groups in the next time window when few groups from timeframe $T_{i+1}$ consist of members of one group from timeframe $T_i$. We can distinguish two types of splitting: (1) equal, which means the contribution of the groups in split group is almost the same and (2) unequal when one of the groups has much greater contribution in the split group, which for this one group the event might be similar to shrinking.

5. Merging (reverse to splitting) – the group has been created by merging several other groups when one group from timeframe $T_{i+1}$ consist of two or more groups from the previous timeframe $T_i$. Merge, just like the split, might be (1) equal, which means the contribution of the groups in merged group is almost the same, or (2) unequal, when one of the groups has much greater contribution into the merged group.

6. Dissolving happens when a group ends its life and does not occur in the next time window, i.e. its members have vanished or stop communicating with each other and scattered among the rest of the groups.

7. Forming (opposed to dissolving) of new group occurs when group which has not existed in the previous time window $T_i$ appears in next time window $T_{i+1}$.

The examples of events described above are presented in Figure 1.

## 2.3. GED – a method for group evolution discovery in the social network

To track social community evolution in social network the new method called GED (Group Evolution Discovery) was developed [3]. Key element of this method is a new measure called inclusion. This measure allows to evaluate the inclusion of one group in another. Therefore, inclusion of group $G_1$ in group $G_2$ is calculated as follows:

$$I(G_1, G_2) = \overbrace{\frac{|G_1 \cap G_2|}{|G_1|}}^{group\ quantity} \cdot \underbrace{\frac{\sum_{x \in (G_1 \cap G_2)} NI_{G_1}(x)}{\sum_{x \in (G_1)} NI_{G_1}(x)}}_{group\ quality} \quad (2)$$

where $NI_{G_1}(x)$ is the value reflecting importance of the node x in group $G_1$.

As a node importance measure, any metric which indicate member position within the community can be used, e.g. centrality degree, betweenness degree, page rank, social position etc. The second factor in Equation 2 would have to be adapted accordingly to selected measure.

As mentioned earlier the GED method, used to track group evolution, takes into account both the quantity and quality of the group members. The quantity is reflected by the first part of the inclusion measure, i.e. what portion of $G_1$ members is shared by both groups $G_1$ and $G_2$, whereas the quality is expressed by the second part of the inclusion measure, namely what contribution of important members of $G_1$ is shared by both groups $G_1$ and $G_2$. It provides a balance between the groups, which contain many of the less important members and groups with only few but key members.

One might say that inclusion measure is "unfair" for not identical groups, because if the community differs even by only one member, inclusion is reduced through not having all nodes and also through not having social position of those nodes. Indeed, it is slightly "unfair" (or rather strict), but using member position within the community calculated on the basis of users relations, makes *inclusion* to focus not only on nodes (members) but also on edges (relations) giving great advantage over methods, which are using only members' overlapping for event identification (group quantity factor in inclusion measure).

It is assumed that only one event may occur between two groups ($G_1$, $G_2$) in the consecutive timeframes, however one group in timeframe $T_i$ may have several events with different groups in $T_{i+1}$.

### *GED* – Group Evolution Discovery Method

**Input:** *TSN* in which at each timeframe $T_i$ groups are extracted by any community detection algorithm. Calculated any user importance measure.
1. For each pair of groups <$G_1$, $G_2$> in consecutive timeframes $T_i$ and $T_{i+1}$ inclusion of $G_1$ in $G_2$ and $G_2$ in $G_1$ is counted according to equation (2).
2. Based on inclusion and size of two groups one type of event may be assigned:
    a. *Continuing*: $I(G_1,G_2) \geq \alpha$ and $I(G_2,G_1) \geq \beta$ and $|G_1| = |G_2|$
    b. *Shrinking*: $I(G_1,G_2) \geq \alpha$ and $I(G_2,G_1) \geq \beta$ and $|G_1| > |G_2|$ OR $I(G_1) < \alpha$ and $I(G_2,G_1) \geq \beta$ and $|G_1| \geq |G_2|$ and there is only one match (matching event) between $G_2$ and all groups in the previous time window $T_i$
    c. *Growing*: $I(G_1,G_2) \geq \alpha$ and $I(G_2,G_1) \geq \beta$ and $|G_1|<|G_2|$ OR $I(G_1,G_2) \geq \alpha$ and $I(G_2,G_1) <\beta$ and $|G_1| \leq |G_2|$ and there is only one match (matching event) between $G_1$ and all groups in the next time window $T_{i+1}$
    d. *Splitting*: $I(G_1,G_2) \geq \alpha$ and $I(G_2,G_1) < \beta$ and $|G_1| \geq |G_2|$ and there is more than one match (matching event) between $G_1$ and all groups in the next time window $T_{i+1}$ OR
    $I(G_1,G_2) < \alpha$ and $I(G_2,G_1) \geq \beta$ and $|G_1| \geq |G_2|$ and there is more than one match (matching event) between $G_1$ and all groups in the next time window $T_{i+1}$
    e. *Merging*: $I(G_1,G_2) \geq \alpha$ and $I(G_2,G_1) < \beta$ and $|G_1| \leq |G_2|$ and there is more than one match (matching event) between $G_2$ and all groups in the previous time window $T_i$ OR
    $I(G_1,G_2) < \alpha$ and $I(G_2,G_1) \geq \beta$ and $|G_1| \leq |G_2|$ and there is more than one match (matching event) between $G_2$ and all groups in the previous time window $T_i$
    f. *Dissolving*: for $G_1$ in $T_i$ and each group $G_2$ in $T_{i+1}$ $I(G_1) < 10\%$ and $I(G_2) < 10\%$
    g. *Forming*: for $G_2$ in $T_{i+1}$ and each group $G_1$ in $T_i$ $I(G_1) < 10\%$ and $I(G_2)< 10\%$

The indicators $\alpha$ and $\beta$ are the GED method parameters which can be used to adjust the method to particular social network and community detection method. After the experiments in [3] authors suggest that the values of $\alpha$ and $\beta$ should be from range [50%; 100%]

Based on the list of extracted events, which have been extracted by GED method for selected group between each two successive timeframes, the group evolution is created. In the example presented in Figure 1 the network consists from eight time windows. The group forms in $T_2$, then by gaining new nodes grows in $T_3$, next splits into two groups in $T_4$, then by losing one node the bigger group is shrinking in $T_5$, both groups continue over $T_6$, next both groups merges with the third group in $T_7$, and finally the group dissolves in $T_8$.

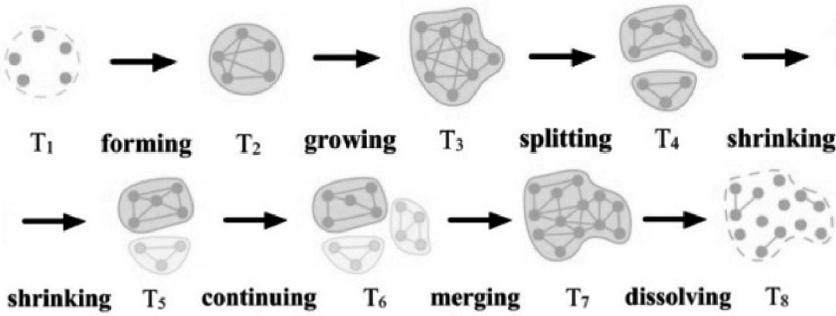

Figure 2.  Changes over time for the single group.

## 3.  User Importance Measures

### 3.1.  Degree Centrality

A degree centrality [9] is the simplest and the most intuitive measure among all. It is the number of links that directly connect one node with others. In an undirected graph it is the number of edges which are connected with the single node. In a directed graph, degree is divided in indegree for edges which are directed to the given node and outdegree for edges which are directed from the given node. The measure is simple, easy to compute and quite informative in many applications [9], but it is local measure and produces big number of duplicates, what is undesirable if we want to create users ranking.

### 3.2.  Social Position

Social position measure can be used to calculate the importance of every single member of the network. The importance of a user described by social position depends on the social positions of first level neighbour and the strength of relationship between user and the neighbours. More precisely user's social position is inherited from neighbours which activity is directed to this user. The social position for the network $SN(V,E)$ is calculated in the iterative way, as follows:

$$SP_{n+1}(x) = (1-\varepsilon) + \varepsilon \cdot \sum_{y \in V} SP_n(y) \cdot C(y \to x), \quad (3)$$

where $SP_{n+1}(x)$ and $SP_n(x)$ is the social position of member $x$ after the $n+1^{st}$ and $n^{th}$ iteration, respectively, and $SP_0(x)=1$ for each $x \in V$; $\varepsilon$ is the fixed coefficient from the range (0;1); $C(y \to x)$ is the commitment function, which expresses the strength of the relation from $y$ to $x$ – the weight of edge $<y,x>$.

## 4. Experiments

Data utilized in the experiments were obtained at the Wroclaw University of Technology. The whole data set was collected within period of February 2006 - October 2007 and consists of 5.845 members and 149.344 relations. The members in this case are WrUT employees and the relation is emails exchange between them. In the experiment overlapping types of timeframe were used, i.e. offset in days of the consecutive time windows is shorter than time window size, so the following time window starts before the previous ends, e.g. the first timeframe begins on the 1st day and ends on the 90th day, second begins on the 46th day and ends on the 135th day and so on (timeframe size is 90 days and offset 45 days). For group extraction the CPM clustering method implemented in CFinder (www.http://cfinder.org/) was utilized. The groups were discovered for clique size of 5 nodes and for the directed and weighted social network.

In the experiments GED method was run: (1) with social position measure [9] [10] (it is very similar to page rank) (2) with degree centrality measure and (3) without any measure, in order to investigate influence of the measure on calculations of inclusion values and also on results of the method. The results obtained with degree centrality as a measure of user importance and results derived without any measure are very similar to the results obtained with social position measure, Table 1.

Table 1. **Results of GED method with different user importance measures.**

| Measure | Execution time [min] | Events found | Threshold $\alpha$ | $\beta$ |
|---|---|---|---|---|
| Social Position | 6 | 1470 | 70 | 70 |
| Degree Centrality | 5:55 | 1447 | 70 | 70 |
| No measure | 5:30 | 1483 | 70 | 70 |

Execution time for GED with degree centrality was slightly better than for the GED with social position because degree centrality value is given as a integer, while the type for social position value is float. The degree centrality was not normalized, as the inclusion measure do not require normalized values and since summing the integers is faster than summing floats, the execution time do degree centrality is shorter. Of course the best execution time was for GED without a measure as an effect of less calculations needed to proceed. Although, the number of events found in all three cases is more or less the same, it can be observed that GED without user importance measure found more events than GED with any of the measures. It is a consequence of the inclusion formula (see equation 2) which consists of two parts. The first fraction is always present, whether GED is run with or without user importance measure, but the second one occurs only when an importance measure is used. Therefore when calculating inclusions of two groups with a measure, it is almost always lower than without any measure. The exceptions are groups where inclusions are equal 100% and groups which do not share any nodes (inclusions equals 0%).

And here comes the question: why *GED* uses a measure of user importance, since it is obvious that it will lower the inclusion? The answer already provided in Section 2.3 is this time supported by clear evidences.

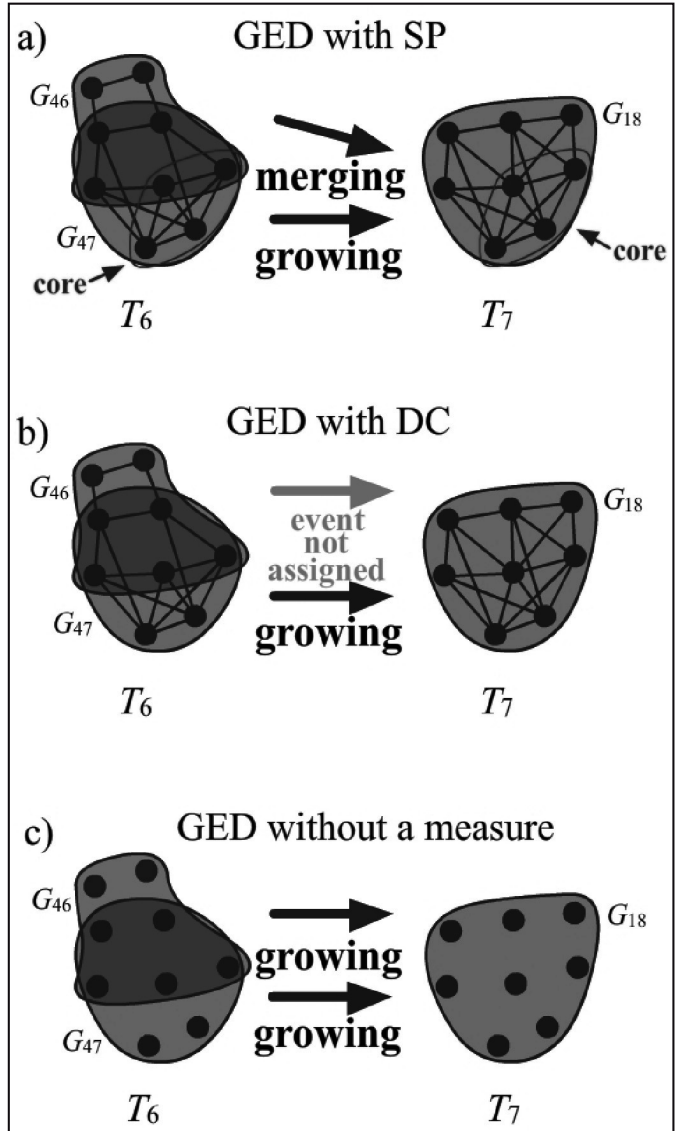

**Figure 3.** Events assigned by GED method with different user importance measures. a) GED with social position measure, arrows marks the core of the group b) GED with degree centrality c) GED without a measure.

As illustrated in Figure 3, two communities $G_{46}$ and $G_{47}$ from time frame $T_6$ overlap by five members and both groups have the same size - seven members. In the next time frame $T_7$ there is only one group $G_{18}$ which consists of all members from the group $G_{47}$ from the previous time frame, and one new member. Two members from the community $G_{46}$ have vanished and are missing in the following time window.

The *GED* method with social position measure, assigned growing event to the community $G_{47}$ and merging event to the group $G_{46}$. The *GED* method with degree centrality measure also assigned growing event to the group $G_{47}$, but did not assign any event to the community $G_{46}$. Finally, *GED* without any user importance measure assigned growing events to both groups from time frame $T_6$.

To have a closer look into the first case, the social positions of members are presented in Table 2. It is clearly visible that the core of the group $G_{47}$ from time frame $T_6$ is identical to the core of the group $G_{18}$ from the next time window $T_7$. The situation is marked with arrow in the Figure 3a and with black dots in the Table 4.5. Additionally, members occurring in the all groups are underlined. Now it is obvious that *GED* with social position measure assigned growing event to group $G_{47}$ because it is almost identical to group $G_{18}$, and "only" merging event to group $G_{46}$ because the cores of both groups have nothing in common. It has to be emphasized once again that, thanks to the user importance measure, *GED* method takes into account both the quantity and quality of the group members providing very accurate results.

Table 2.    Social position of members presented in Figure 7.11a.

| Group | Time window | Node | SP | Rank |
|---|---|---|---|---|
| 46 | 6 | 1443 | 1,48 | 1 |
| 46 | 6 | 3145 | 1,33 | 2 |
| 46 | 6 | <u>7564</u> | 0,96 | 3 |
| 46 | 6 | <u>1326</u> | 0,86 | 4 |
| 46 | 6 | <u>11999</u> | 0,85 | 5 |
| 46 | 6 | <u>14151</u> | 0,77 | 6 |
| 46 | 6 | <u>621</u> | 0,75 | 7 |
| 47 | 6 | 2066• | 1,31 | 1 |
| 47 | 6 | 7328• | 1,30 | 2 |
| 47 | 6 | <u>7564</u>• | 1,28 | 3 |
| 47 | 6 | <u>11999</u>• | 1,04 | 4 |
| 47 | 6 | <u>1326</u> | 0,80 | 5 |
| 47 | 6 | <u>14151</u> | 0,67 | 6 |
| 47 | 6 | <u>621</u> | 0,60 | 7 |
| 18 | 7 | 2066• | 1,49 | 1 |
| 18 | 7 | 7328• | 1,35 | 2 |
| 18 | 7 | <u>7564</u>• | 1,29 | 3 |
| 18 | 7 | <u>11999</u>• | 1,24 | 4 |
| 18 | 7 | <u>1326</u> | 0,75 | 5 |
| 18 | 7 | <u>14151</u> | 0,71 | 6 |
| 18 | 7 | <u>621</u> | 0,66 | 7 |
| 18 | 7 | 4632 | 0,51 | 8 |

The *GED* method with degree centrality measure was even more strict in the studied case, Figure 3b. Low degree centrality within the group $G_{46}$ causes that no event was assigned. In turn, similar structure between groups $G_{47}$ and $G_{18}$ effects in assigning the merging event. Structure of all groups and degree centrality of all members is presented in Table 3. Again, members occurring in all groups are underlined.

Table 3. Degree centrality of members presented in Figure 3a.

| Group | Time window | Node | DC | Rank |
|---|---|---|---|---|
| 46 | 6 | <u>11999</u> | 3 | 1 |
| 46 | 6 | <u>14151</u> | 3 | 1 |
| 46 | 6 | 1443 | 2 | 3 |
| 46 | 6 | 3145 | 2 | 3 |
| 46 | 6 | <u>7564</u> | 2 | 3 |
| 46 | 6 | <u>1326</u> | 2 | 3 |
| 46 | 6 | <u>621</u> | 2 | 3 |
| 47 | 6 | 2066 | 5 | 1 |
| 47 | 6 | 7328 | 5 | 1 |
| 47 | 6 | <u>7564</u> | 4 | 3 |
| 47 | 6 | <u>11999</u> | 4 | 3 |
| 47 | 6 | <u>1326</u> | 4 | 3 |
| 47 | 6 | <u>14151</u> | 3 | 6 |
| 47 | 6 | <u>621</u> | 3 | 6 |
| 18 | 7 | <u>7564</u> | 7 | 1 |
| 18 | 7 | 7328 | 5 | 2 |
| 18 | 7 | 2066 | 5 | 2 |
| 18 | 7 | <u>11999</u> | 5 | 2 |
| 18 | 7 | <u>1326</u> | 5 | 2 |
| 18 | 7 | <u>14151</u> | 4 | 6 |
| 18 | 7 | <u>621</u> | 4 | 6 |
| 18 | 7 | 4632 | 3 | 8 |

Figure 3c presents in the best way how *GED* method without a user importance measure understands the communities. There is no core, all members are equal and relations between them are not considered at all. Such simplification causes that the events assigned to the groups are not the most adequate to situation (but only when comparing with events assigned by *GED* with user importance measure). Having information only about the members in the groups but not about their relations results in incorrect events assignment. Thus, if researchers investigating group evolution are not interested in groups structure and relations between members, a simpler and faster version of the GED Method may be successfully used. However, if there is enough time to calculate any user importance measure, it is recommended to use the *GED* method in the original version.

## 5.  Conclusions

The GED method is very flexible and allows researcher to adjust the method to one's needs. One of the adjustments is the possibility to use this user importance measure which is best suited for the analysed network. For example, degree centrality may be used in the network with very large groups where more accurate global measure like social position or betweenness cannot be used, due to their computational complexity.

In this paper the analysis of three cases was conducted, i.e. (1) global measure - social position,(2) local measure - degree centrality and (3) no user importance measure.

The experiments show how particular user importance measure affect the results of the GED method. While using (2) and (3) is the fastest way to get the results it may lead to not precise results. Of course sometimes, e.g. huge networks with very large groups, one may has to use them but it should be aware of the consequences.

## References

[1] Asur, S., Parthasarathy, S., Ucar, D.: An event-based framework for characterizing the evolutionary behavior of interaction graphs. *ACM Trans. Knowl. Discov. Data.* 3, 4, Article 16, 2009, 36 pages.

[2] Bródka P., Musiał K., Kazienko P.: A Performance of Centrality Calculation in Social Networks. In Proceedings of the 2009 International Conference on Computational Aspects of Social Networks (CASON '09). *IEEE Computer Society*, Washington, DC, USA, 2009, 24-31.

[3] Bródka P., Saganowski P., Kazienko P.: GED: The Method for Group Evolution Discovery in Social Networks, *Social Network Analysis and Mining*, 2012, DOI:10.1007/s13278-012-0058-8

[4] Chakrabarti D, Kumar R, Tomkins A, *Evolutionary Clustering*, KDD 2006, ACM, Philadelphia.

[5] Lin, Y.R., Chi, Y., Zhu, S., Sundaram, H., Tseng, B.L. *FacetNet: A Framework for Analyzing Communities and Their Evolutions in Dynamic Networks*, WWW 2008, ACM, 2008, pp. 685-694.

[6] Musiał K.., Kazienko P., Bródka P.: User position measures in social networks. In Proceedings of the 3rd Workshop on Social Network Mining and Analysis (SNA-KDD '09). *ACM*, New York, NY, USA, , Article 6 , 2009, 9 pages.

[7] Palla, G., Barabási, A.L., Vicsek, T.: Quantifying social group evolution. *Nature* 446, 2007, 664-667.

[8] Sun J, Papadimitriou S, Yu P.S, Faloutsos C, *GraphScope: Parameter-free Mining of Large Time-evolving Graphs*. KDD, ACM, 2007, 687-696.

[9] Wasserman S., Faust K., *Social network analysis: Methods and applications*, New York: Cambridge University Press, 1994.